\documentclass[fleqn,twoside,]{article}
\usepackage{espcrc2}
\input epsf.tex

\usepackage{graphicx}

\newcommand{\gtrsim}{ \mathop{}_{\textstyle \sim}^{\textstyle >} }
\newcommand{\lesssim}{ \mathop{}_{\textstyle \sim}^{\textstyle <} }

\title{Curvaton mechanism and its implications to sneutrino
cosmology
} 

\author{Takeo Moroi\address{%
Department of Physics,  Tohoku University, Sendai 980-8578, Japan}
}
       
\begin{document}

\begin{abstract}

I describe basic features of the curvaton scenario where the
primordial fluctuation of a late-decaying scalar field, called
``curvaton,'' becomes the dominant source of the cosmic density
fluctuations.  I also discuss its implications to sneutrino cosmology.


\vspace{1pc}
\end{abstract}

\maketitle

\renewcommand{\thefootnote}{\#\arabic{footnote}}
\setcounter{footnote}{0}
\renewcommand{\theequation}{\thesection.\arabic{equation}}

\section{Introduction}
\label{sec:intro}
\setcounter{equation}{0}

In the study of the evolution of the universe, it is important to
understand the origin of the cosmic density fluctuations.  The most
conventional scenario is inflation \cite{inflation} where quantum
fluctuation of the inflaton field during the inflation becomes the
origin of the cosmic density fluctuations.  From the particle-physics
point of view, it is desirable to construct a model of inflation which
is testable by collider (or other laboratory) experiments; if it is
possible, we can study the mechanism of the cosmological density
fluctuations with collider experiments.  It is, however, difficult to
construct such a testable model of inflation since the requirements on
the inflaton potential is very stringent.\footnote
{Within the minimal supersymmetric standard model, however, it may be
possible to use some of the scalar quarks and Higgs bosons as the
inflaton.  For detail, see \cite{Kasuya:2003iv}.}

Recently, a new mechanism of generating the cosmic density
fluctuations has been attracting many attentions, where a
late-decaying scalar condensation provides the dominant source of the
cosmic density fluctuations.  In this scenario, dominant part of the
cosmic density fluctuations originate from the primordial fluctuation
of a new scalar field, called ``curvaton,'' which is different from
the inflaton field \cite{Enqvist:2001zp,Lyth:2001nq,Moroi:2001ct}.
Even though, in a large class of the curvaton scenario, inflation is
assumed as a solution to the horizon and flatness problems as well as
to generate the primordial fluctuation of the curvaton field,
constraints on the inflaton potential can be relaxed in the curvaton
scenario.  In addition, since the requirements on the curvaton is not
so stringent, it is possible to use (some of the) well-motivated
particles as the curvaton; for example, scalar fields in the minimal
supersymmetric standard model (MSSM) may play the role of the
curvaton.  In such a case, it may be possible to study the properties
of the fields responsible for the structure formation by collider
experiments.  (For detailed studies of the curvaton scenario, see
\cite{RecentCurvaton}.)

Here, I will review the curvaton scenario and discuss some of its
implications to particle cosmology.  Organization of the rest of this
article is as follows.  In Section \ref{sec:history}, I first discuss
the thermal history we consider.  Then, in Section
\ref{sec:mechanism}, I explain the curvaton scenario.  The cosmic
microwave background (CMB) angular power spectrum generated in the
curvaton scenario is studied in Section \ref{sec:cl}.  In Section
\ref{sec:implications}, implications to the sneutrino cosmology is
discussed.  Section \ref{sec:conclusion} is devoted for conclusions
and discussion.

\section{Thermal history}
\label{sec:history}
\setcounter{equation}{0}

I first discuss the basic scenario.  In the curvaton scenario, there
are two scalar fields which play important roles; one is the inflation
$\chi$ and the other is the curvaton field $\phi$.\footnote
{I assume the inflation to solve the horizon and flatness problems.
The curvaton mechanism may be implemented in the pre-big-bang
\cite{pbb} and the ekpyrotic \cite{ekpyrotic} scenarios.  In this
article, however, I will not discuss those cases.}
The most important features of the curvaton scenario do not depend on
the details of the inflation.  Thus, I do not assume any specific form
of the inflaton potential.  On the contrary, resultant density
fluctuations depend on the potential of the curvaton field.  Here, for
simplicity, I adopt the simplest form of the curvaton potential, i.e.,
the parabolic one:
\begin{eqnarray}
V(\phi) = \frac{1}{2} m_\phi^2 \phi^2,
\end{eqnarray}
where $m_\phi$ is smaller than the expansion rate of the universe
during the inflation $H_{\rm inf}$.  In addition, the curvaton field
is assumed to have non-vanishing initial amplitude $\phi_{\rm init}$.

In the scenario, the universe starts with the inflationary epoch and,
after the inflation, the universe is reheated by the decay of the
inflaton.  Then, the universe is dominated by the radiation (which I
call $\gamma_\chi$).  I call the radiation-dominated epoch filled with
radiation generated from the decay products of the inflaton as the
first radiation dominated epoch, or RD1 epoch, since, in the curvaton
scenario, there are two radiation-dominated epochs.  Denoting the
decay rate of the inflaton as $\Gamma_\chi$, reheating temperature
after the inflation is given by\footnote
{In the following discussions, I neglect $O(1)$ coefficients which are
not important.}
\begin{eqnarray}
T_{\rm R1} \sim \sqrt{M_*\Gamma_\chi},
\label{T_R1}
\end{eqnarray}
where $M_*$ is the reduced Planck scale.  In the early stage of the
RD1 epoch, slow-roll condition is satisfied for the curvaton field.
As the universe expands, however, the curvaton field starts to
oscillate.  When $\Gamma_\phi\lesssim H\lesssim m_\phi$ (with
$\Gamma_\phi$ being the decay rate of $\phi$), energy density of
$\phi$ behaves as that of non-relativistic matter.  Then, the energy
density of the curvaton $\rho_\phi$ is proportional to $a^{-3}$ while
the energy density of $\gamma_\chi$ is proportional to $a^{-4}$, where
$a$ is the scale factor.  Thus, as the universe expands, curvaton
dominates the universe (if the lifetime of the curvaton is long
enough).  I call this epoch as curvaton-dominated epoch, or $\phi$D
epoch.  The curvaton field decays when the expansion rate becomes
comparable to the decay rate of the curvaton.  The reheating
temperature due to the curvaton decay is given by
\begin{eqnarray}
T_{\rm R2} \sim \sqrt{M_*\Gamma_\phi}.
\label{T_R2}
\end{eqnarray}

\section{Curvaton mechanism: evolutions of the fluctuations}
\label{sec:mechanism}
\setcounter{equation}{0}

In the curvaton scenario, $\phi$ also acquires the primordial
fluctuation during the inflation.  Denoting the curvaton field (with
comoving momentum coordinate $\vec{x}$) as\footnote
{Here, I  use the same notation for the zero-mode and for the total
amplitude since there should be no confusion.}
\begin{eqnarray}
\phi (t, \vec{x}) = \phi (t) + \delta\phi (t, \vec{x}),
\end{eqnarray}
let us consider the two-point correlation function of the fluctuation:
\begin{eqnarray}
    &&
    \langle 0 | \delta\phi (t, \vec{x}) \delta\phi (t, \vec{y}) 
    | 0 \rangle =
    \nonumber \\ && ~~~
    \int \frac{d k}{k} \frac{d\Omega_{\vec{k}}}{4\pi} 
    |\delta\phi (t, \vec{k})|^2
    e^{i \vec{k} (\vec{x}-\vec{y})}.
\end{eqnarray}
Then, the Fourier amplitude generated during the inflation is given by
\begin{eqnarray}
    \delta\phi (t, \vec{k}) = 
    \left( \frac{k}{2aH_{\rm inf}}
    \right)^{m_\phi^2/3H_{\rm inf}^2}
    \left[ \frac{H_{\rm inf}}{2\pi} \right]_{k=aH_{\rm inf}},
    \nonumber \\
    \label{dphi_init}
\end{eqnarray}
where the subscript ``$k=aH_{\rm inf}$'' implies that the quantity is
evaluated at the time of the horizon-exit during the inflation.  As I
discuss below, the primordial fluctuation of $\phi$ given in Eq.\
(\ref{dphi_init}) becomes the origin of the cosmic density fluctuations
in the curvaton scenario.

Now, we are at the position to discuss the evolutions of the
cosmological density fluctuations generated from $\delta\phi_{\rm
init}$.  For this purpose, I use the Newtonian gauge where the line
element is described with the metric perturbations $\Psi$ and $\Phi$
as\footnote
{Here, I use the notation and convention of \cite{Hu:1995em} unless
otherwise mentioned.}
\begin{eqnarray}
    ds^2 &~=~& - (1 + 2\Psi) dt^2
    + a^2 (1 + 2\Phi)  d\vec{x}^2 
    \nonumber \\  &~=~&
    a^2 
    \left[ - (1 + 2\Psi) d\tau^2 + (1 + 2\Phi) d\vec{x}^2 \right],
    \label{ds}
\end{eqnarray}
where $\tau$ is the conformal time.  In addition, the variable
$\delta_X$ is defined as
\begin{eqnarray}
\delta_X \equiv \delta \rho_X / \rho_X,
\end{eqnarray}
where the subscript $X$ denotes the individual components (like
radiation, cold dark matter (CDM), baryon, and so on) and
$\delta\rho_X$ is the fluctuation of the energy density of $X$.

Substituting Eq.\ (\ref{ds}) into the Einstein equation, we obtain the
generalized Poisson equation for $\Phi$:
\begin{eqnarray}
    k^2 \Phi &~=~& 
    \frac{1}{2M_*^2} a^2  
    \rho_{\rm tot} 
    \left[ 
        \delta_{\rm tot} + \frac{3 \mathcal{H}}{k} 
        (1+\omega_{\rm tot}) V_{\rm tot}
    \right].
    \nonumber \\
    \label{poisson}
\end{eqnarray}
Here, ``tot'' denotes the total matter and the variable $V_X$ denotes
the velocity perturbation of the component $X$.  Furthermore,
$\omega_{\rm tot}\equiv\rho_{\rm tot}/p_{\rm tot}$ is the
equation-of-state parameter for the total matter, and
\begin{eqnarray}
    \mathcal{H} \equiv \frac{1}{a} \frac{da}{d\tau}.
\end{eqnarray}
I consider the situation where the temperature of the universe is so
high that the momentum-exchange of relativistic particles are
efficient enough.  In this case, perturbation of the radiation becomes
locally isotropic and the anisotropic stress perturbation vanishes,
resulting in $\Psi +\Phi=0$.

When the perturbation of the radiation becomes locally isotropic, the
equations for the density and velocity perturbations of the radiation
are given by
\begin{eqnarray}
    \delta_\gamma' &~=~& -\frac{4}{3} k V_\gamma - 4 \Phi', 
    \label{dr'} \\
    V_\gamma' &~=~& \frac{1}{4} k \delta_\gamma + k\Psi,
    \label{Vr'}
\end{eqnarray}
where the ``prime'' denotes the derivative with respect to the
conformal time $\tau$.  In addition, if a very weakly interacting
non-relativistic component exists, its perturbations obey the
following equations:
\begin{eqnarray}
    \delta_m' &~=~& -k V_m - 3\Phi',
    \label{eq:cdm} \\
    V_m'  &~=~& - \mathcal{H} V_m + k \Psi,
    \label{eq:cdm_velocity}
\end{eqnarray}
where the subscript ``$m$'' is for non-relativistic matters.  Notice
that, when a scalar field is oscillating, the equation-of-state
parameter of the scalar condensation vanishes and hence the density
and velocity perturbations of the scalar field also obey Eqs.\ 
(\ref{eq:cdm}) and (\ref{eq:cdm_velocity}).

In the curvaton scenario, it is assumed that the primordial
fluctuation $\delta\phi_{\rm init}$ becomes the dominant source of the
cosmic density fluctuations.  Thus, hereafter, I will discuss the
density fluctuations generated from $\delta\phi_{\rm init}$.  When
$\Gamma_\phi\lesssim H\lesssim m_\phi$, the situation is like the case
with isocurvature fluctuation in non-relativistic matter component
\cite{Mollerach:hu}.  In the curvaton case, as in the conventional
isocurvature case, it is convenient to define the (primordial) entropy
fluctuation between $\phi$ (and its decay products) and the photon
from the inflaton $\gamma_\chi$:
\begin{eqnarray}
S_{\phi\chi} \equiv 
\left[ \delta_\phi - \frac{3}{4} \delta_{\gamma_\chi} 
\right]_{\rm RD1}.
\end{eqnarray}
Then, the cosmological density fluctuations are parameterized by using
$S_{\phi\chi}$.

In order to discuss the evolutions of the fluctuations, it is
important to know the equation-of-state parameters of individual
components in the universe.  If all the components behave as the
relativistic or the non-relativistic matter, evolutions of the
perturbations are described by Eqs.\ (\ref{dr'}) $-$
(\ref{eq:cdm_velocity}).  In this case, it is convenient to
distinguish the photon (or any other components) from the decay
product of $\phi$ from that from the inflaton field, which I call
$\gamma_\phi$ and $\gamma_\chi$, respectively.\footnote
{In fact, these photons are mixed each other and they cannot be
defined separately.  In other words, their velocity perturbations
should be the same since they form a single fluid.  Even so, the
following arguments are unchanged as far as we consider the leading
terms in the density perturbations since the velocity perturbation is
at most $O(k\tau)$.  In the following discussion, $\gamma_\phi$ and
$\gamma_\chi$ should be understood as representatives of the
components which are and are not generated from the decay product of
$\phi$, respectively.}
In order to consider $\delta_{\gamma_\phi}$ in the RD2 epoch, we can
neglect $\gamma_\chi$ since the CMB radiation at this epoch is
dominantly from the $\phi$ field.  Then, we find that, in the RD2
epoch, $\Psi$ and $\delta_{\gamma_\phi}$ become constant while
$V_{\gamma_\phi}=O(k\tau)$ up to higher order corrections.  Indeed,
combining Eq.\ (\ref{poisson}) with Eqs.\ (\ref{dr'}) and (\ref{Vr'}),
and using $\delta_{\rm tot}=\delta_{\gamma_\phi}$ and $V_{\rm
tot}=V_{\gamma_\phi}$, we obtain
$V_{\gamma_\phi}^{(\delta\phi)}=-\frac{1}{2}k\tau\Psi_{\rm
RD2}^{(\delta\phi)}$ and
\begin{eqnarray}
    \delta_{\gamma_\phi}^{(\delta\phi)} 
    = -2 \Psi_{\rm RD2}^{(\delta\phi)},
\end{eqnarray}
where $\Psi_{\rm RD2}^{(\delta\phi)}$ is the metric perturbation
induced by the primordial fluctuation of the amplitude of $\phi$.  (In
the following, the superscript ``$(\delta\phi)$'' is for perturbations
generated from the primordial fluctuation of $\phi$.)  As I mentioned,
$\Psi_{\rm RD2}^{(\delta\phi)}$ is constant up to a correction of
$O(k^2\tau^2)$.

Behavior of $\delta_{\gamma_\chi}^{(\delta\phi)}$ is also easily
understood.  In discussing the effects of the primordial fluctuation
of $\phi$, we neglect the initial fluctuation of the inflaton field
and hence $\delta_{\gamma_\chi}^{(\delta\phi)}\rightarrow 0$ in the
deep RD1 epoch.  In addition, from Eqs.\ (\ref{dr'}) and (\ref{Vr'}),
$V_{\gamma_\chi}^{(\delta\phi)}$ becomes higher order in $k\tau$ than
$\delta_{\gamma_\chi}^{(\delta\phi)}$ and $\Psi^{(\delta\phi)}$.
Thus, we obtain
\begin{eqnarray}
    \delta_{\gamma_\chi}^{(\delta\phi)} = 4 \Psi^{(\delta\phi)}.
\end{eqnarray}
The above relation holds in the RD1, $\phi$D, and RD2 epochs up to
corrections of $O(k^2\tau^2)$.

$[\Psi^{(\delta\phi)}]_{\rm RD2}$ can be related to $S_{\phi\chi}$.
Using the fact that the entropy fluctuation is a conserved quantity
for superhorizon mode, the following relation holds:
\begin{eqnarray}
    S_{\phi\chi}^{(\delta\phi)} &~=~& 
    \left[ \frac{3}{4} \delta_{\gamma_\phi}^{(\delta\phi)}
        - \frac{3}{4} \delta_{\gamma_\chi}^{(\delta\phi)}
    \right]_{\rm RD2}
    \nonumber \\  &~=~&
    \left[ \delta_{\phi}^{(\delta\phi)}
        - \frac{3}{4} \delta_{\gamma_\chi}^{(\delta\phi)}
    \right]_{\rm \phi D, RD1}.
\end{eqnarray}
With this relation, in particular, we obtain
\begin{eqnarray}
\Psi_{\rm RD2}^{(\delta\phi)} = 
- \frac{2}{9} S_{\phi\chi}^{(\delta\phi)}.
\end{eqnarray}

Density fluctuations of various components are also parameterized by
$S_{\phi\chi}$.  Detailed properties of the density fluctuations,
however, depend on how the various components in the universe are
produced.  If a component $X$ is generated from the decay product of
$\phi$, then there is no entropy between the photon (i.e.,
$\gamma_\phi$) and $X$.  On the contrary, if some other scalar field
$\psi$ generates $X$, the entropy between the photon and $X$ is the
same as $S_{\phi\chi}$.  Thus, if all the components in the universe
are generated from $\phi$, the density fluctuations become purely
adiabatic and
\begin{eqnarray}
    \left[ \delta_\gamma^{(\delta\phi)} \right]_{\rm RD2} 
    &~=~& \frac{4}{3}\left[ \delta_b^{(\delta\phi)} \right]_{\rm RD2}
    \nonumber \\
    &~=~& \frac{4}{3}\left[ \delta_c^{(\delta\phi)} \right]_{\rm RD2}
    \nonumber \\
    &~=~& -2 \Psi_{\rm RD2}^{(\delta\phi)},
\end{eqnarray}
where the subscripts $\gamma$, $b$, and $c$ are for the photon,
baryon, and CDM, respectively.  In this case, the isocurvature
perturbation in the $\phi$ field is converted to the purely adiabatic
density perturbation after the decay of $\phi$.  On the contrary, if
the baryon asymmetry is generated by the scalar field $\psi$, the
entropy between the radiation and the baryon becomes
$S_{\phi\chi}^{(\delta\phi)}$ and hence \cite{Moroi:2001ct,MorTak}
\begin{eqnarray}
    &&
    \left[ \delta_\gamma^{(\delta\phi)} \right]_{\rm RD2} 
    = \frac{4}{3}\left[ \delta_c^{(\delta\phi)} \right]_{\rm RD2}
    = -2 \Psi_{\rm RD2}^{(\delta\phi)},
    \\ &&
    \left[ \delta_b^{(\delta\phi)} \right]_{\rm RD2}
    = \frac{3}{4} \left[ \delta_\gamma^{(\delta\phi)} \right]_{\rm RD2} 
    + \frac{9}{2} \Psi_{\rm RD2}^{(\delta\phi)},
    \label{Sb}
\end{eqnarray}
and in the case where $\psi$ is responsible for the CDM while the
baryon number is somehow generated from the decay product of $\phi$,
\begin{eqnarray}
    &&
    \left[ \delta_\gamma^{(\delta\phi)} \right]_{\rm RD2} 
    = \frac{4}{3}\left[ \delta_b^{(\delta\phi)} \right]_{\rm RD2}
    = -2 \Psi_{\rm RD2}^{(\delta\phi)},
    \\ &&
    \left[ \delta_c^{(\delta\phi)} \right]_{\rm RD2}
    = \frac{3}{4} \left[ \delta_\gamma^{(\delta\phi)} \right]_{\rm RD2} 
    + \frac{9}{2} \Psi_{\rm RD2}^{(\delta\phi)}.
    \label{Sc}
\end{eqnarray}
In addition, if the baryon and the CDM are both generated from sources
other than $\phi$, we obtain
\begin{eqnarray}
    &
    \left[ \delta_\gamma^{(\delta\phi)} \right]_{\rm RD2}
    & =
    -2 \Psi_{\rm RD2}^{(\delta\phi)},
    \\ &
    \left[ \delta_b^{(\delta\phi)} \right]_{\rm RD2}
    & =
    \left[ \delta_c^{(\delta\phi)} \right]_{\rm RD2}
    \nonumber \\ &&
    = \frac{3}{4} \left[ \delta_\gamma^{(\delta\phi)} \right]_{\rm RD2} 
    + \frac{9}{2} \Psi_{\rm RD2}^{(\delta\phi)}.
    \label{Sbc}
\end{eqnarray}
It is important to notice that, for the cases given in Eqs.\ 
(\ref{Sb}) $-$ (\ref{Sbc}), the isocurvature perturbation is
correlated with the adiabatic perturbation.

So far, we have seen that the primordial fluctuation of the curvaton
field may generate the metric and density fluctuations.  Before
closing this section, we should compare the size of the curvaton
contribution with the inflaton contribution.  Even with the curvaton,
there is also a contribution from the inflaton fluctuation since I
assume inflation as a solution to the horizon and flatness problems.
If we consider the situation where there is no entropy fluctuation,
inflaton contribution to the metric perturbation (in the RD2 epoch) is
given by \cite{Bardeen:qw}
\begin{eqnarray}
\Psi^{\rm (inf)}_{\rm RD2} (k) =
\frac{2}{3}
\left[ 
\frac{3H_{\rm inf}^2}{V'_{\rm inf}} \times 
\frac{H_{\rm inf}}{2\pi} 
\right]_{k=aH},
\label{Psi(inf)}
\end{eqnarray}
with $V'_{\rm inf}\equiv\partial V_{\rm inf}/\partial\chi$, while the
curvaton contribution is
\begin{eqnarray}
\Psi^{(\delta\phi)}_{\rm RD2} (k) =
-\frac{4}{9}
\left[ 
\frac{1}{\phi_{\rm init}} \times
\frac{H_{\rm inf}}{2\pi} 
\right]_{k=aH}.
\label{Psi(curv)}
\end{eqnarray}
(Here and hereafter, the superscript ``(inf)'' is for quantities
generated from the inflaton fluctuation.)  As one can see, the
curvaton contribution to $\Psi$ is inversely proportional to
$\phi_{\rm init}$ and hence, if $\phi_{\rm init}$ is small enough,
curvaton contribution dominates over the inflaton contribution.

\begin{figure}
    \centerline{\epsfxsize=0.5\textwidth\epsfbox{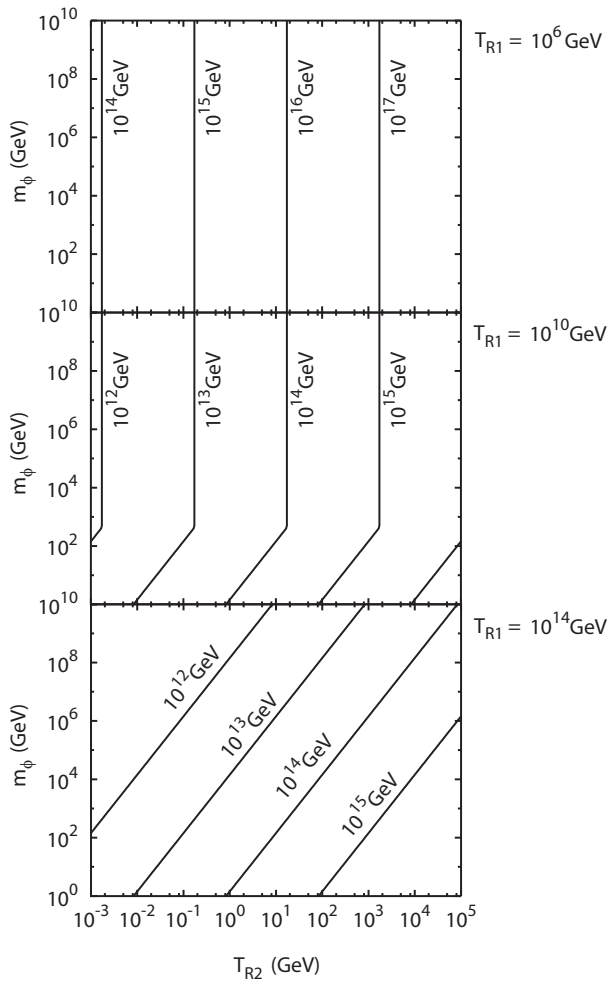}}
    \caption{Lower bound on $\phi_{\rm init}$ to realize the $\phi$D
    epoch on $T_{\rm R2}$ vs.\ $m_\phi$ plane.  $T_{\rm R1}$ is taken
    to be $10^{6}$GeV, $10^{10}$GeV and $10^{14}$GeV from the top.}
    \label{fig:phimin}
\end{figure}

Thus, in the curvaton scenario, there are two conflicting requirements
on the initial amplitude of the curvaton field.  One is the upper
bound on $\phi_{\rm init}$; upper bound on $\phi_{\rm init}$ for
$\Psi^{(\delta\phi)}\gtrsim\Psi^{\rm (inf)}$ depends on the model of
the inflation.  For the case of the chaotic inflation, for example,
the curvaton contribution becomes larger than the inflaton
contribution if $\phi_{\rm init}\lesssim M_*$.  The other is the lower
bound on the $\phi_{\rm init}$, since the $\phi$D epoch cannot be
realized if $\phi_{\rm init}$ is too small.  Lower bound depends on
the reheating temperatures $T_{\rm R1}$ and $T_{\rm R2}$ given in
Eqs.\ (\ref{T_R1}) and (\ref{T_R2}), respectively.  In Fig.\ 
\ref{fig:phimin}, I plot the lower bound on $\phi_{\rm init}$ to
realize the $\phi$D epoch.

\section{CMB angular power spectrum}
\label{sec:cl}
\setcounter{equation}{0}

Now, I discuss the CMB angular power spectrum in the curvaton
scenario.  One of the most important consequences of the curvaton
scenario is that, if all the components in the universe are generated
only from the decay products of $\phi$, no entropy fluctuation is
generated and the primordial density fluctuations (after the RD2
epoch) becomes purely adiabatic.  Let us first consider such a case.

One important check point is that the scale dependence of the
primordial density fluctuations.  In the case where the cosmic density
fluctuations are generated from the primordial fluctuation of the
inflaton, scale-dependence originates from the change of the slope of
the inflaton potential as well as the expansion rate during the
inflation.  As can be seen from Eq.\ (\ref{Psi(curv)}), on the
contrary, scale-dependence of the curvaton contribution is only from
the change of the expansion rate (as far as $m_\phi\ll H_{\rm inf}$).
Thus, even though both the curvaton and inflation contributions are
from primordial fluctuations of some scalar fields, their
scale-dependences are different.  Defining the spectral index $n_{\rm
S}$ as
\begin{eqnarray}
\frac{d}{d\ln k}
\ln [\Psi^{({\rm inf}, \delta\phi)}_{\rm RD2}]^2 
\simeq n^{({\rm inf}, \delta\phi)}_{\rm s}-1,
\end{eqnarray}
the spectral indices for the inflaton and curvaton contributions are
calculated as
\begin{eqnarray}
n^{\rm (inf)}_{\rm s} &~=~& 1 - 6 \epsilon + 2 \eta,
\\
n^{(\delta\phi)}_{\rm s} &~=~& 1 - 2 \epsilon,
\end{eqnarray}
where $\epsilon$ and $\eta$ are slow-roll parameters which are given
by
\begin{eqnarray*}
\epsilon = 
\frac{1}{2} M_*^2 
\left( \frac{V_{\rm inf}'}{V_{\rm inf}} \right)^2,~~~
\eta = M_*^2 \frac{V_{\rm inf}''}{V_{\rm inf}}.
\end{eqnarray*}

Due to the change of the scale-dependence, we can see that the
observational constraints on inflation models are relaxed in the
curvaton scenario.  For example, for the case of the chaotic inflation
with the inflaton potential $V_{\rm inf}\propto \chi^{p_{\rm inf}}$,
spectral indices for the inflaton and curvaton contributions are
estimated as $n^{\rm (inf)}_{\rm s}\simeq 0.96$ and
$n^{(\delta\phi)}_{\rm s}\simeq 0.98$ for $p_{\rm inf}=2$, $n^{\rm
(inf)}_{\rm s}\simeq 0.95$ and $n^{(\delta\phi)}_{\rm s}\simeq 0.97$
for $p_{\rm inf}=4$, and $n^{\rm (inf)}_{\rm s}\simeq 0.94$ and
$n^{(\delta\phi)}_{\rm s}\simeq 0.95$ for $p_{\rm inf}=6$.  Thus,
using currently available constraint on the spectral index $n_{\rm
s}=0.99\pm 0.04$ \cite{Spergel:2003cb} from the Wilkinson Microwave
Anisotropy Probe (WMAP), simple chaotic inflation model with $p_{\rm
inf}=6$ is excluded by the observations while, with the curvaton
mechanism, $p_{\rm inf}=6$ becomes consistent with the observational
constraints.  For other class of inflation models, change of the
constraint may be more drastic.

Next, let us discuss the effects of the possible entropy fluctuation
in the curvaton scenario.  As mentioned in the previous section, in
the curvaton scenario, correlated mixture of the adiabatic and
isocurvature fluctuations may be generated.  In order to discuss
effects of the entropy fluctuation to the CMB angular power spectrum,
it is convenient to parameterize the density fluctuation of the
non-relativistic matter as
\begin{eqnarray}
    \left[ \delta_m^{(\delta\phi)} \right]_{\rm RD2} &~\equiv~&
    \left[ \frac{\Omega_b}{\Omega_m} \delta_b^{(\delta\phi)} 
        + \frac{\Omega_c}{\Omega_m} \delta_c^{(\delta\phi)} 
    \right]_{\rm RD2}
    \nonumber \\ &~=~&
    \frac{3}{4} \left[ \delta_\gamma^{(\delta\phi)} \right]_{\rm RD2} 
    + \kappa_m \Psi_{\rm RD2}^{(\delta\phi)},
\end{eqnarray}
where $\Omega_b$, $\Omega_c$ and $\Omega_m$ are the (present) density
parameters of the baryon, the CDM, and the non-relativistic component
(and hence $\Omega_m=\Omega_b+\Omega_c$).  Although the density
fluctuations for the baryon and the CDM are independent, shape of the
CMB angular power spectrum is determined once $\kappa_m$ is fixed.
For the purely adiabatic case, $\kappa_m=0$.  With the relations
(\ref{Sb}), (\ref{Sc}), and (\ref{Sbc}), $\kappa_m$ becomes
$\frac{9}{2}(\Omega_b/\Omega_m)$, $\frac{9}{2}(\Omega_c/\Omega_m)$,
and $\frac{9}{2}$, respectively.

\begin{figure}[t]
    \begin{center}
    \centerline{\epsfxsize=0.5\textwidth\epsfbox{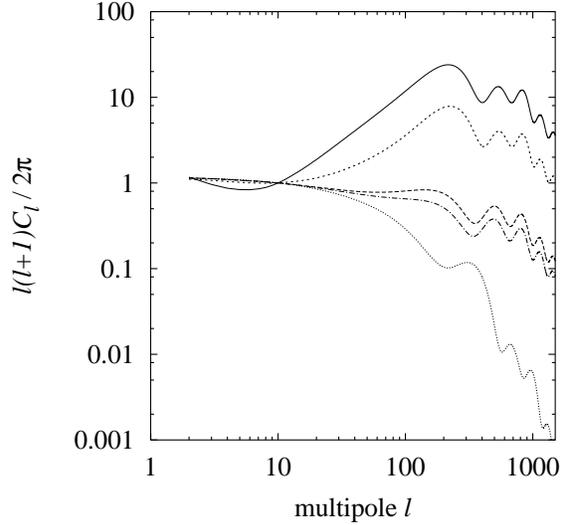}}
    \caption{The angular power spectrum with correlated mixture of the
    adiabatic and isocurvature perturbations in the baryonic sector
    (solid line), in the CDM sector (long-dashed line), and in the
    baryonic and CDM sectors (dot-dashed line).  (See Eqs.\
    (\ref{Sb}), (\ref{Sc}), and (\ref{Sbc}), respectively.)  I also
    show the CMB angular power spectrum for the purely adiabatic
    (i.e., $\kappa_m=0$) and purely isocurvature ($\kappa_m=\infty$)
    cases with short-dashed and dotted lines, respectively.  The
    overall normalizations are taken as $[l(l+1)C_l/2\pi]_{l=10}=1$.}
    \label{fig:clkdep}
    \end{center}
\end{figure}

In Fig.\ \ref{fig:clkdep}, I show how the CMB angular power spectrum
depends on the $\kappa_m$ parameter.  As one can see, $C_l$ strongly
depends on $\kappa_m$ and, if $|\kappa_m|\gtrsim O(0.1)$, deviation of
the CMB angular power spectrum from the adiabatic result (i.e., $C_l$
with $\kappa_m=0$) becomes sizable.  Importantly, the WMAP results
strongly suggest that the primordial density fluctuations be (almost)
purely adiabatic; with the WMAP data, $\kappa_m< 0.1$ is obtained
\cite{HamKawMorTak}.  Thus, the WMAP results impose stringent
constraint on the curvaton scenario.  A possible small contamination
of the entropy fluctuation in some case will be discussed in the next
section.

\section{Implications to sneutrino cosmology}
\label{sec:implications}
\setcounter{equation}{0}

Now, I would like to discuss the case where the curvaton field is also
responsible for the scenario of baryogenesis.  Among various
scenarios, there are some cases where the baryon asymmetry of the
universe is generated from scalar-field condensations.  In those
scenarios, scalar fields often dominate the universe at some epoch and
hence, if they have primordial fluctuations, they may play the role of
the curvaton.  Probably, two of the most famous examples of the
possible scalar fields responsible for the baryon asymmetry are
right-handed sneutrino \cite{Murayama:1992ua,Hamaguchi:2001gw} and the
Affleck-Dine field \cite{Affleck:1984fy}.\footnote
{In some case, inflaton can be responsible for the scenario of the
baryogenesis.  For example, right-handed neutrinos can be produced by
the decay of the inflaton, which may become the origin of the baryon
asymmetry of the universe \cite{Asaka:1999jb}.}

Here, I consider the case where the right-handed sneutrino, which
becomes the origin of the baryon asymmetry of the universe
\cite{Murayama:1992ua}, plays the role of the curvaton \cite{MorMur}.
First, I briefly summarize the model and the thermal history.  Here,
the relevant part of the superpotential is given by
\begin{eqnarray}
W = \hat{h}_{N,i\alpha} N_i L_\alpha H_u + 
\frac{1}{2} \hat{M}_{N,ij} N_i N_j,
\end{eqnarray}
where $\hat{h}_N$ is the Yukawa matrix for the neutrino while
$\hat{M}_N$ is the Majorana mass matrix for the right-handed
(s)neutrinos.  Here, $i$ and $j$ are generation indices of the
right-handed neutrino $N$ while $\alpha$ is that of the left-handed
lepton doublet $L$.  In addition, $H_u$ is the up-type Higgs field.  I
work in the basis where the matrix $\hat{M}_N$ is diagonalized.  For
simplicity, let us consider the case where the lightest right-handed
sneutrino $\tilde{N}=\tilde{N}_1$ has non-vanishing initial amplitude.
(Hereafter, mass of $\tilde{N}$ is denoted as $M_N$.)

With a non-vanishing primordial amplitude, $\tilde{N}$ starts to
oscillate when $H\sim M_N$ and decays when $H\sim \Gamma_N$, where
$\Gamma_N$ is the decay rate of $\tilde{N}$.  Since the Majorana mass
term breaks the lepton-number symmetry, lepton-number asymmetry may be
generated at the time of the sneutrino decay if non-vanishing CP
violation exists.  Such a lepton-number asymmetry becomes the source
of the baryon-number asymmetry of the universe with the sphaleron
process.

The mechanism of generating the baryon-number asymmetry is basically
the supersymmetric version of the Fukugita-Yanagida mechanism
\cite{Fukugita:1986hr}.  Expression for the resultant amount of the
baryon asymmetry is, however, rather complicated since, in this case,
the primordial abundance of the right-handed sneutrino has
non-thermally determined.  Assuming $\Gamma_N<\Gamma_\chi$, we obtain
\cite{Hamaguchi:2001gw}
\begin{eqnarray}
    \frac{n_B}{s} &~\simeq~& 
    0.24 \times 10^{-10} 
    \nonumber \\ && \times
    f_{\gamma_{\tilde{N}}} \delta_{\rm eff}
    \left(
        \frac{T_{N}}{10^6 {\rm GeV}}
    \right)
    \left(
        \frac{m_{\nu_3}}{0.05 {\rm eV}}
    \right),
    \label{nB/s(N-dom)}
\end{eqnarray}
where $T_{N}$ is the temperature at the epoch of the decay of
$\tilde{N}$, $m_{\nu_3}$ the mass of the heaviest (left-handed)
neutrino mass.  (So, if $\tilde{N}$ decays after dominating the
universe, $T_{N}$ becomes the reheat temperature due to the decay of
$\tilde{N}$.)  In addition, in the basis where the Majorana mass
matrix for the right-handed neutrinos $\hat{M}$ is real and
diagonalized, the effective CP violating phase is given by
\begin{eqnarray}
    \delta_{\rm eff} = \frac{\langle H_u\rangle^2}{m_{\nu_3}}
    \frac{ {\rm Im} 
    [ \hat{h}\hat{h}^\dagger \hat{M}^{-1}\hat{h}^*\hat{h}^T ]_{11} }
    { [\hat{h} \hat{h}^\dagger]_{11} }.
\end{eqnarray}
Notice that, with maximum CP violation, $\delta_{\rm eff}\sim 1$.  In
addition, $f_\gamma$ is the energy fraction of the radiation generated
from the decay product of $\tilde{N}$.  With the relation
$\Gamma_N<\Gamma_\chi$, $\tilde{N}$ decays after the inflaton decay.
In this case, it is convenient to define the following quantity:
\begin{eqnarray}
    \tilde{N}_{\rm eq} \sim 
    \left\{
        \begin{array}{ll}
            (\Gamma_N / M_N)^{1/4} M_* & 
            ~:~ M_N < \Gamma_\chi \\
            (\Gamma_N /\Gamma_\chi)^{1/4} M_* &
            ~:~ M_N > \Gamma_\chi
        \end{array}
    \right. .
\nonumber \\
\end{eqnarray}
If $\tilde{N}_{\rm init}\sim \tilde{N}_{\rm eq}$,
$\rho_{\gamma_\chi}\sim\rho_{\tilde{N}}$ is realized when
$H\sim\Gamma_{\tilde{N}}$.  Thus, when $\tilde{N}_{\rm init}\lesssim
\tilde{N}_{\rm eq}$, $\tilde{N}$ decays in the $\gamma_\chi$-dominated
universe and hence
\begin{eqnarray}
    f_{\gamma_{\tilde{N}}}\sim
    \frac
    {(\tilde{N}_{\rm init}/\tilde{N}_{\rm eq})^2}
    {1+(\tilde{N}_{\rm init}/\tilde{N}_{\rm eq})^2}
    ~:~ 
    \tilde{N}_{\rm init} \lesssim \tilde{N}_{\rm eq}.
    \label{f(init<eq)}
\end{eqnarray}
On the contrary, if $\tilde{N}_{\rm init}\gtrsim \tilde{N}_{\rm eq}$,
the right-handed sneutrino decays after it dominates the universe and
we obtain
\begin{eqnarray}
    f_{\gamma_{\tilde{N}}}\sim
    \frac
    {(\tilde{N}_{\rm init}/\tilde{N}_{\rm eq})^{8/3}}
    {1+(\tilde{N}_{\rm init}/\tilde{N}_{\rm eq})^{8/3}}
    ~:~ 
    \tilde{N}_{\rm eq} \lesssim \tilde{N}_{\rm init}.
    \label{f(eq<init)}
\end{eqnarray}
Thus, as is easily seen, energy fraction of $\gamma_{\tilde{N}}$
becomes close to 1 when $\tilde{N}_{\rm init}\gg\tilde{N}_{\rm eq}$
while, in the opposite limit, most of the radiation are generated from
the decay product of the inflaton.

Now, I consider the cosmological density fluctuations in this
scenario. In particular, I concentrate on the case where the
condensation of the right-handed sneutrino plays the role of the
curvaton.  (Thus, in this case, the epoch after the decay of
$\tilde{N}$ is identified as the RD2 epoch.)  The scenario is
basically the same as the curvaton scenario which I have explained
before.

In the previous discussion, I have discussed the cases where all the
components in the universe are generated from the decay product of the
curvaton, which corresponds to the case of
$f_{\gamma_{\tilde{N}}}\simeq 1$.  As is explained above, however,
(small) contamination of the radiation from the decay product of the
inflaton may always exist.  If the baryon asymmetry is generated from
thermally produced particles after the RD2 epoch is realized, then
entropy fluctuations vanish and the primordial density fluctuations
become purely adiabatic.

In the scenario we consider, however, the situation is quite different
since the baryon asymmetry is generated from the decay product of
$\tilde{N}$.  In particular, if the energy fraction of the radiation
from the decay product of the inflaton becomes sizable, non-vanishing
baryonic entropy fluctuation may be generated.

Evolution of the density fluctuations in this case is also understood
by solving the Einstein and Boltzmann equations for the fluctuations
given in the previous section.  We can calculate the entropy in the
radiation-dominated universe after the decay of both $\chi$ and
$\tilde{N}$.  In the radiation-dominated universe \cite{Hu:1995em}
\begin{eqnarray}
\Delta_{\rm tot} &~=~& O(k^2 \tau^2),
\\
\delta_{\rm tot} &~=~& -2 \Psi_{\rm RD2},
\\
V_{\rm tot} &~=~& \frac{1}{2} \Psi_{\rm RD2} k \tau.
\end{eqnarray}
Then, with the relation
\begin{eqnarray}
\Delta_{\rm tot} = f_{\gamma_{\chi}} \Delta_{\gamma_{\chi}} +
f_{\gamma_{\tilde{N}}} \Delta_{\gamma_{\tilde{N}}},
\end{eqnarray}
with $\Delta_{\gamma_X}=\delta_{\gamma_X}+4 (a'/a)V_{\rm tot}/k$, we
obtain
\begin{eqnarray}
\Delta_{\gamma_{\tilde{N}}}^{(\delta\tilde{N})} =
- \frac{f_{\gamma_{\chi}}}{f_{\gamma_{\tilde{N}}}}
\Delta_{\gamma_{\chi}}^{(\delta\tilde{N})} =
- 6\frac{f_{\gamma_{\chi}}}{f_{\gamma_{\tilde{N}}}}
\Psi_{\rm RD2}^{(\delta\tilde{N})}.
\end{eqnarray}
Using the fact that the entropy between any component produced from
$\tilde{N}$ and that from the inflaton field is conserved, we can
relate $\Psi_{\rm RD2}^{(\delta\tilde{N})}$ with
$S_{\tilde{N}\chi}^{(\delta\tilde{N})}$; with the relation
$S_{\tilde{N}\chi}^{(\delta\tilde{N})}
=\frac{3}{4}(\Delta_{\gamma_{\tilde{N}}}^{(\delta\tilde{N})} -
\Delta_{\gamma_{\chi}}^{(\delta\tilde{N})})$, we obtain
\begin{eqnarray}
\Psi_{\rm RD2}^{(\delta\tilde{N})} = 
-\frac{2}{9} f_{\gamma_{\tilde{N}}} 
S_{\tilde{N}\chi}^{(\delta\tilde{N})}
= -\frac{4}{9} f_{\gamma_{\tilde{N}}} 
\frac{\delta\tilde{N}_{\rm init}}{\tilde{N}_{\rm init}}.
\end{eqnarray}
Thus, if $\tilde{N}$ decays much after it dominates the universe,
$f_{\gamma_{\tilde{N}}}\simeq 1$ and hence the metric perturbation
becomes comparable to the primordial entropy perturbation.  On the
other hand, if $f_{\gamma_{\tilde{N}}}\ll 1$, the metric perturbation
becomes negligibly small.
Since the baryon asymmetry is generated from $\tilde{N}$, the density
fluctuation in the baryonic component is given by
\begin{eqnarray}
\Delta_b^{(\delta\tilde{N})} = 
\frac{3}{4} \Delta_{\gamma_{\tilde{N}}}^{(\delta\tilde{N})} = 
- \frac{9}{2} \frac{f_{\gamma_{\chi}}}
{f_{\gamma_{\tilde{N}}}} \Psi_{\rm RD2}^{(\delta\tilde{N})}.
\end{eqnarray}
Thus, the entropy between the baryon and the radiation is given by
\begin{eqnarray}
S_{b\gamma}^{(\delta\tilde{N})} &~=~& 
\Delta_b^{(\delta\tilde{N})} - 
\frac{3}{4} \Delta_{\rm tot}^{(\delta\tilde{N})}
\nonumber \\
&~=~& - \frac{9}{2} 
\frac{f_{\gamma_{\chi}}}{f_{\gamma_{\tilde{N}}}} 
\Psi_{\rm RD2}^{(\delta\tilde{N})}
\nonumber \\
&~=~& -
\frac{9(1 - f_{\gamma_{\tilde{N}}})}
{2f_{\gamma_{\tilde{N}}}} 
\Psi_{\rm RD2}^{(\delta\tilde{N})},
\label{S_bg}
\end{eqnarray}
and defining
\begin{eqnarray}
\kappa_b^{(\delta\tilde{N})} \equiv
\frac{S_{b\gamma}^{(\delta\tilde{N})}}
{\Psi_{\rm RD2}^{(\delta\tilde{N})}},
\end{eqnarray}
we obtain
\begin{eqnarray}
\kappa_b^{(\delta\tilde{N})} = 
-\frac{9(1 - f_{\gamma_{\tilde{N}}})}
{2f_{\gamma_{\tilde{N}}}}.
\end{eqnarray}
If the right-handed sneutrino once dominates the universe,
$f_{\gamma_{\tilde{N}}}\rightarrow 1$ and hence the perturbation
becomes adiabatic.  (Thus, the situation is like the case discussed in
the previous section.)

It is interesting if a sizable amount of the energy density of the
radiation is from the inflaton.  In this case,
$f_{\gamma_{\tilde{N}}}$ becomes smaller than $1$ and correlated
mixture of the adiabatic and isocurvature fluctuations is generated.
As mentioned in the previous section, WMAP data suggests that the
baryonic and CDM entropy fluctuations should be very small;
$|\kappa_b|<0.5$ \cite{HamKawMorTak}.  If $f_{\gamma_{\tilde{N}}}\sim
0.1$, however, $|\kappa_b|$ can be smaller than the present upper
bound but the deviation of the CMB angular power spectrum from the
adiabatic result may still be within the reach of the future precise
observations.  Thus, it is desirable to find the signal from the
(correlated) entropy fluctuation in the future precise observations of
the CMB anisotropy for the test of the curvaton scenario.

Finally, let me comment on the case where the curvaton also generates
the baryon asymmetry of the universe via Affleck-Dine mechanism.  If
the primordial fluctuation of the Affleck-Dine field becomes the
dominant source of the cosmological density fluctuations, large
baryonic entropy fluctuation is generated even if the universe is once
completely dominated by the Affleck-Dine field.  This is due to the
fact that the Affleck-Dine field is a complex scalar field and also
that the resultant baryon asymmetry of the universe is sensitive to the
initial value of the Affleck-Dine field.  If the Affleck-Dine field
acquires primordial fluctuation, its initial phase as well as the
initial amplitude is expected to have fluctuations.

\section{Conclusions and discussion}
\label{sec:conclusion}
\setcounter{equation}{0}

Here, I have discussed the basic features of the curvaton scenario and
its implications to particle cosmology.  In particular, I have
discussed that the cosmic density fluctuations can dominantly
originate from the primordial fluctuation of the ``curvaton'' field,
which is a late-decaying scalar condensation other than the inflaton.
In the curvaton scenario, scale-dependence of the cosmic density
fluctuation becomes different from the case of the simple inflation,
and in many cases, the spectral index $n_{\rm s}$ becomes close to $1$
compared to the simple inflation case.  Thus, given the fact that the
CMB anisotropy observed by the WMAP suggests (almost) scale-invariant
primordial density fluctuations, observational constraints on the
inflaton potential can be relaxed by adopting the curvaton scenario.

One of the important feature of the curvaton scenario is the possible
contamination of the (correlated) entropy fluctuation in the
non-relativistic matter.  The CMB angular power spectrum is sensitive
to the entropy fluctuation.  Thus, if such correlated entropy
fluctuation is sizable, it affects the shape of the CMB angular power
spectrum and hence it will be interesting and important to try to find
its consequence in particular in the CMB anisotropy at future
precision observations of the universe.

\section*{Acknowledgments}

The author is grateful to the organizers of the Fujihara Seminar
``SEESAW 1979 -- 2004: Neutrino Mass and Seesaw Mechanism'' for their
hospitality during the stay.  The work of T.M. is supported by the
Grant-in-Aid for Scientific Research from the Ministry of Education,
Science, Sports, and Culture of Japan, No.\ 15540247.

\end{document}